\documentstyle[aaspp4,psfig]{article}

\begin{document}

\title{The Spectrum of GRB 930131 (``Superbowl Burst'') \\
from 20 keV to 200 MeV }

\author{Volker Bromm}
\affil{Department of Astronomy, Yale University, New Haven, CT 06520-8101;
volker@astro.yale.edu}

\and
\author{Bradley E. Schaefer }
\affil{Department of Physics, Yale University, New Haven, CT 06520-8121;
schaefer@grb2.physics.yale.edu}

\begin{abstract}

We have constructed a broad-band spectrum for GRB 930131
(the ``Superbowl Burst''), ranging from 20 keV to 200 MeV, by
combining spectral information from the {\it Gamma Ray Observatory}'s BATSE, COMPTEL
and EGRET instruments. We present general methods for combining spectra
from different time intervals obtained by the same instrument as well
as for combining spectra from the same time interval taken by different
instruments. The resulting spectrum is remarkably
flat (in $\nu F_{\nu}$-space) up to high energies. We
find that the spectral shape can be successfully fitted
by the shocked synchrotron emission model of Tavani. We present evidence
that the flatness of the spectrum at high energies is not due to
spectral time-variability.

\end{abstract}

\section{INTRODUCTION}

 Broad-band spectra of gamma-ray bursts (GRBs)
pose a difficult challenge to any theoretical model trying to
explain them. Looking only at a limited range of energy, as, for example, each
of the different instruments on board the {\it Gamma Ray Observatory} (GRO) 
does individually, results in
a featureless power law perhaps with some curvature. However, a broad-band spectrum, ranging over
many decades in energy, typically contains interesting features like
peaks, curvature and breaks. Such features will be diagnostic of the
physical processes in the burst fireball and the spectra can be used
to directly test models of burst emission. Only a few broad-band spectra
have been produced (Schaefer et al. 1998; Greiner et al. 1995; Hurley
et al. 1994), as only bright bursts detected by multiple instruments 
on board the GRO have a wide enough range of available data. The
brightest such burst is GRB 930131 which reached a peak flux of
$105 \mbox{\ ph} \mbox{\ s}^{-1} \mbox{\ cm}^{-2}$ (Meegan et al. 1996). This burst has
BATSE trigger number 2151 and has been called the ``Superbowl Burst'' after
its time of occurence. EGRET and COMPTEL spectra have already appeared
in the literature (Sommer et al. 1994; Ryan
et al. 1994), but no BATSE spectrum has been presented due to
severe deadtime problems. 

This paper is organized in the following way. In \S 2, we provide general
methods for combining spectra obtained by the same instrument during
different time intervals (2.1.), 
as well as for combining spectra taken by different instruments covering the
same time interval (2.2.).
These methods can
be used in many common GRB applications, provided the necessary requirements
are met. In \S 3, we carry out the construction of the broad-band spectrum
of GRB 930131 from 20 keV to 200 MeV.
First, we describe how the individual (BATSE, COMPTEL, and
EGRET) spectra have been obtained (3.1.).
Then, we argue why these independently reduced spectra
can be combined with the method of \S 2, where we point out
the non-obliging nature of this procedure in the present case. After
presenting the resulting spectrum (3.2.), we compare this to theoretical
models of the GRB emission mechanism (3.3.). Subsequently, we discuss
evidence for spectral evolution (3.4.). Finally, \S 4 summarizes the
spectral properties of this remarkable burst.

\section{COMBINING SPECTRA}
Often the problem occurs to combine individual spectra into either a
time-averaged or an instrument-averaged spectrum. The second case arises
in cross-calibrating spectral information from instruments that are
sensitive in different energy ranges.
In this section, it is  
assumed that the observed count spectra have already been reduced into
photon spectra. 
In the following, we describe the method of combining spectra and
give the relevant formulae, which are then applied to the case of
GRB~930131 in Section 3.

\subsection{Combining Across Time}
Suppose the time over which one wants to average is divided up into
smaller time intervals $k$ with respective livetimes $\tau_{k}$. For
each time interval $k$ and energy bin $i$ the photon flux
(in units of photons/area/energy/time) is
$\left(\frac{dn}{de}\right)_{ik}$ with standard deviation $\sigma_{ik}$.
Then, constructing the time-averaged spectrum is straightforward. With
the total livetime given by 
$\tau_{total}=\sum_{k}\tau_{k}$, the time-averaged photon flux in 
energy bin $i$ is
\begin{equation}
\left(\frac{dN}{dE}\right)_{i}=\tau^{-1}_{total}\sum_{k}\left
(\frac{dn}{de}\right)_{ik}
\tau_{k} \mbox{\ \ ,}
\end{equation}
and the resulting standard deviation is
\begin{equation}
\sigma_{i}^{2}=\tau^{-1}_{total}\sqrt{\sum_{k}\left(\sigma_{ik}\tau_{k}
\right)^{2}}
\mbox{\ \ .}
\end{equation}

\subsection{Combining Across Different Instruments}
Spectra from different instruments can be combined just as can spectra
from multiple detectors on the same instrument. We here assume that the
combination process is robust, i.e., that the resulting spectrum is not
greatly obliging (cf., Section 3.2.). 
This has to be justified on a case by case basis.
Another requirement
is that either the input spectra are for identical time
intervals, or they cover the entire burst. This
combination can be described as a four-step process:

{\it Step A}:\\
The spectra from different instruments are divided into energy bins in
different ways. Therefore, as a first step, all the bin boundaries 
($E^{low}$ and $E^{high}$)
from all the instruments are put into increasing order and then used
to define subbins. Assume that after the ordering, the following
sequence arises:\, 
$\mbox{...}<E_{k-1}<E_{k}<E_{k+1}<\mbox{...}$\,
Then define the $k$th subbin to cover an energy interval between $E_{k}$ and
$E_{k+1}$. Figure 1 illustrates this procedure for the case of two
instruments.

{\it Step B}:\\
It is preferable to conduct the combining in $\nu F_{\nu}$-space, 
where $\nu F_{\nu}\propto\left(\frac{dN}{dE}\right)E^{2}$. Then the 
spectrum is roughly constant over a given energy bin, as opposed
to the usual steep decline in ordinary $\frac{dN}{dE}$-space. Now, for
energy bin $i$ of instrument $m$,
having lower and higher energies $E_{mi}^{low}$ and $E_{mi}^{high}$,\,
respectively, define the energy flux 
per logarithmic energy interval
\begin{equation}
\left(\frac{d\varphi}{dE}\right)_{mi}\equiv\left(\frac{dN}{dE}\right)_{mi}\left(E_{mi}^{mid}\right)^{2} \pm \sigma_{mi}\mbox{\ \ ,} 
\end{equation}
where $E_{mi}^{mid}=\sqrt{E_{mi}^{low}\cdot E_{mi}^{high}}$ and
$\sigma_{mi}$ is the uncertainty of $\left(\frac{d\varphi}{dE}\right)_{mi}$.
Our procedure presumes that $\left(\frac{d\varphi}{d\varphi}\right)_{mi}$
changes little across each energy bin, as is the case for energy bins
that are small compared to either the detector resolution or the structure
in the spectrum. This covers
virtually all GRB applications, although a simple interpolation scheme
might be appropriate for a particularly steep spectrum observed with
very broad bins.

{\it Step C}:\\
Now, we want to cross-combine the
spectra of different instruments.
In constructing the spectrum for subbin $k$, we first determine whether
a given instrument $m$ has an energy bin $i$ overlapping the subbin.
If this is the case, we set
\begin{equation}
\left(\frac{d\varphi}{dE}\right)_{mk}=\left(\frac{d\varphi}{dE}\right)_{mi} 
\mbox{\ \ \ \ \ and\ \ \ \ \ } \sigma_{mk}=\sigma_{mi}\mbox{\ \ .} 
\end{equation}
Figure 1 shows the case of two instruments having overlapping energy
bins with subbin $k$. The energy flux
of the cross-combined spectrum is the weighted average of
all contributing spectra:
\begin{equation}
\left(\frac{d\phi}{dE}\right)_{k}  = \sigma_{k}^{2} \cdot \sum_{m}
\frac{1}{\sigma_{mk}^{2}}\left(\frac{d\varphi}{dE}\right)_{mk}\mbox{\ \ ,}
\end{equation}
where
\begin{equation}
\sigma_{k}=\left(\sum_{m}\sigma_{mk}^{-2}\right)^{-1/2}\mbox{\ \ .}
\end{equation}

{\it Step D}:\\
As a last step, put together the subbins into larger bins of width
appropriate for the spectral resolution and features.
Rebinning, e.g., two subbins $k$ and $k+1$ into a larger bin $l$
with boundaries $E_{l}^{low}$ and $E_{l}^{high}$ is accomplished by the following:
\begin{equation}
\left(\frac{d\Phi}{dE}\right)_{l}=w_{k} \left(\frac{d\phi}{dE}\right)_{k}
 + w_{k+1} \left(\frac{d\phi}{dE}\right)_{k+1}\mbox{\ \ ,} 
\end{equation}
where one has for the respective weights
\begin{equation}
w_{k}=
\frac{E_{k+1}-E^{low}_{l}}{E^{high}_{l}-E^{low}_{l}}\mbox{\ \ ,} 
\end{equation}
and
\begin{equation}
w_{k+1}=
\frac{E^{high}_{l}-E_{k+1}}{E^{high}_{l}-E^{low}_{l}}\mbox{\ \ .} 
\end{equation}
The resulting standard deviation is
\begin{equation}
\sigma^{2}_{l}=w_{k}^{2}\sigma^{2}_{k}
 + w_{k+1}^{2}\sigma_{k+1}^{2}\mbox{\ \ .}
\end{equation}
If the output bin covers more than two subbins, then equations (7)-(10) can be
easily generalized or used repeatedly.
\placefigure{fig1}

\section{THE SPECTRUM OF GRB 930131}
\subsection{The Individual Spectra}
For all 3 instruments (BATSE, EGRET, COMPTEL), their photon spectra
have been obtained by the traditional forward-folding technique 
(Loredo \& Epstein 1989). This technique assumes a variety of spectral
models $M$, and convolves them with the respective detector response
matrix (DRM), symbolically $C_{model}=\mbox{DRM}\ast M$, where $C_{model}$ is 
the count spectrum predicted by the model. The parameters of the model are then
adjusted to obtain the best fit to the observed count spectrum,
$C_{obs}=\mbox{DRM}\ast P_{true}$, where $P_{true}$ is the true (photon)
spectrum of the source. Alternatively, a model-independent inverse technique
could have been adopted, where $P_{true}=\mbox{DRM}^{-1}\ast C_{obs}$.
Attempts at doing so have proven unconvincing, and the nearly universal
practice in gamma-ray astronomy is to use forward-folding techniques.
One exception is the direct inversion 
method of Pendleton et al. (1996), which has only been applied to low
resolution (4-channel) data and introduces considerable additional
error (10-15\%).

\subsubsection{\it BATSE Spectrum}
The ``Superbowl Burst'' suffers from severe deadtime effects,\,which is the
reason why the original discovery paper (Kouveliotou et al.\,1994) does
not present a spectrum for the BATSE energy range. For this bright burst, most of
the flux arrives in the first 0.06~seconds, a situation which saturates the
BATSE Large Area Detectors (LADs), whereas the smaller but thicker Spectroscopy
Detectors (SDs) can reliably record the intense photon flux. In constructing our
spectrum, we have selected the two burst-facing Spectroscopy Detectors (SD 4 and 5), for
which there are available the well suited
STTE-data (SD Time-Tagged Events), which cover
the first $\sim$~1.5~s of the burst and which have a time resolution
of $128\mbox{\,} \mu$s. Therefore, we can correct for the deadtime effects by
subdividing the total time into 53 individual spectra with a duration
of as short as a few ms around the first, intense peak. For each time interval, 
the photon spectrum is obtained by following the procedure described in
Schaefer et al. (1994). In carrying out the forward-folding, we assume a
single power-law spectral model.
Then, by applying the methods of
Section 2.1., we constructed time-averaged spectra for SD 4 and 5
, which were then in turn combined (as described in Section 2.2.) to
give the overall spectrum for the BATSE 
energy-range (21 keV to 1.18 MeV, above which the flux-errors
exceed 100\%).

\subsubsection{\it COMPTEL And EGRET Spectra}
The COMPTEL and EGRET spectra have previously been published (Ryan et al. 1994,
and Sommer et al. 1994, respectively) and we refer the reader to these
papers for details. We have chosen to work with the spectrum reported
by the EGRET Total Absorption Shower Counter (TASC), since the EGRET spark
chamber is too severely affected by deadtime effects. The TASC spectrum
covers an energy range from 1 MeV to 180 MeV. 
The overlap region between BATSE and
TASC is nicely covered by the COMPTEL instrument, where the COMPTEL Telescope
spectrum covers the range from 0.75 Mev to 30 MeV. Both spectra have
been obtained by the forward-folding technique with a power-law model, and
are corrected for deadtime effects.

\subsection{The Combined Spectrum}
To construct the combined spectrum with the method described
in Section 2.2., we first have to ascertain the robustness of this procedure.
It is well known (Fenimore et al. 1983) that the resulting spectral shape
can possibly depend sensitively on the details
of the fitting technique (i.e., that
the spectra might be ``obliging''). In principle, it could make a big
difference whether the low- and high-energy parts, covered by different
instruments, are first unfolded separately and only then combined together, or
whether the unfolding is done simultaneously to all instruments. The
physical reason for this is that high-energy photons might masquerade
as low-energy ones, and that, consequently, the low-energy part of
the spectrum cannot be accurately
unfolded independently of the high-energy part.
For the present case, however, this problem does not occur. It has
been convincingly shown that the BATSE Spectroscopy Detectors are
non-obliging (Schaefer et al. 1994; cf., their Figures 11 and 52).
This is primarily due to their thickness, which largely minimizes photon
energies being underreported.
The TASC and COMPTEL spectra, on the other
hand, are not affected by the lower energy BATSE range. Finally, treating
the COMPTEL and TASC spectra independently of each other is rendered
possible by the fact that the model fitting leads to almost
identical results ($\frac{dN}{dE}\propto E^{-2}$). We are
therefore justified in combining the independently obtained spectra
from the 3 GRO
instruments (BATSE-EGRET-COMPTEL) into the overall, broad-band spectrum
of GRB 930131.

This combination is carried out with
the method of section 2.2., where we have been careful
to construct our BATSE spectrum such that it exactly matches the
time coverage of the EGRET TASC instrument, and approximately that
of COMPTEL. To evaluate how well the instruments agree in the mutual
overlap region around 1 MeV, we compare the fluxes at 1 MeV for the
3 instruments (in units of $10^{-3}$photons cm$^{-2}$ sec$^{-1}$ keV$^{-1}$): 
BATSE 2$\pm$2, 
COMPTEL 8$\pm$3, and TASC $2\pm$0.5. The agreement between BATSE and TASC
is good, although the BATSE errors approach
100\% at these high energies. The COMPTEL flux is
somewhat high, but due to its uncertainties it does not
contribute significantly to the weighted average of
the final, combined spectrum.

Table 1 and Figure 2 present
the $\nu F_{\nu}$ ($\propto \left(\frac{dN}{dE}\right)E^{2}$) spectrum
in units of (photons s$^{-1}$ cm$^{-2}$ keV$^{-1}$)$\ast (E^{mid}/100
\mbox{\ keV})^{2}$.
The resulting spectrum is remarkably flat, as compared
to other published broad-band spectra, which have a much more peaked
appearance (cf., Schaefer et al. 1998). In the following section we ask, whether
this rather unusual spectral shape is consistent with the model of
shocked synchrotron emission, which successfully fits the characteristics
of other broad-band GRB spectra.
Subsequently, we investigate
whether the flat spectrum of GRB 930131 can be understood as a result of
spectral evolution.

\placefigure{fig2}
\placetable{spec}

\subsection{Model-Fits}
We fit our combined spectrum to the shocked synchrotron model
of Tavani (1996a, b),\,which gives the following analytical expression
for the energy flux:
\begin{equation}
\psi_{model}\equiv\left(\frac{d\Phi}{dE}\right)=\nu F_{\nu}=C\nu\left[I_{1} + \frac{1}{e}I_{2}\right]
\end{equation}
\begin{equation}
I_{1}=\int_{0}^{1}y^{2}e^{-y}F\left(\frac{\nu}{\nu_{c}^{\ast}y^{2}}\right)\mbox{d}y
\end{equation}
\begin{equation}
I_{2}=\int_{1}^{\infty}y^{-\delta}F\left(\frac{\nu}{\nu_{c}^{\ast}y^{2}}\right)\mbox{d}y\mbox{\ \ ,}
\end{equation}
where $F(x)\equiv x\int_{x}^{\infty}\mbox{K}_{\frac{5}{3}}(w)\mbox{d}w$ is
the usual synchrotron spectral function with $\mbox{K}_{\frac{5}{3}}$ being
the modified Bessel-function of order $\frac{5}{3}$ and $e=2.718...$.\,The
normalization constant $C$ has units of specific flux.\,
Equations
(12) and (13) are summing up the synchrotron emission from a Maxwellian
distribution of electron energies which breaks to a power law at high energies.\,
Here, $\delta$ is the index of the supra-thermal power-law distribution of 
particles, resulting from relativistic shock-acceleration. The critical
frequency $\nu_{c}^{*}$ describes where most of the synchrotron power is emitted.
We apply the Levenberg-Marquardt method of non-linear $\chi^{2}$ fitting
(cf., Numerical Recipes, Press et al. 1992) to minimize 
\begin{equation}
\chi^{2}=\sum_{i=1}^{N}\left(\frac{\psi_{i}-\psi_{model}(\nu_{i}^{mid};C,\delta,\nu_{c}^{\ast})}
{\sigma_{i}}\right)^{2}\mbox{\ \ .}
\end{equation}
Our observed spectrum with flux $\psi_{i}=\left(\frac{d\Phi}{dE}\right)_{i}$ and uncertainty $\sigma_{i}$ contains $N=37$ data points.
\,Our best-fit parameters are: 
\begin{equation}
C=104 \pm 8 \mbox{\ \ erg cm$^{-2}$ sec$^{-1}$ Hz$^{-1}$}
\end{equation}
\begin{equation}
\delta=3.3 \pm 0.1
\end{equation}
\begin{equation}
h\nu_{c}^{\ast}=98 \pm 14 \mbox{\ \ keV}
\end{equation}
The fit has a chi-squared of $\chi^{2}=38$ with 34 degrees of freedom.
Therefore, we can conclude that the spectrum of
GRB 930131 is consistent with the Tavani-model. At low 
energies,\,the spectrum is asymptotically approaching 
$\nu F_{\nu} \propto \nu^{4/3}$, as is usual for burst spectra (Schaefer
et al. 1998). This behavior 
is predicted by optically thin synchrotron theory (Katz 1994).

\subsection{Spectral Evolution}
All of the published broad-band spectra (Schaefer et al. 1998; Greiner 
et al. 1995; Hurley et al. 1994) are
strongly peaked and fall off steeply above the peak energy. GRB~930131, on
the other hand, has a spectrum which remains constant (within a factor of 4)
over four orders of magnitude in energy. Can this
behavior be understood as the result of a superposition of many spectra, which
individually show the usual, strongly peaked shape and whose peak energy
evolves with time? For the BATSE energy range, the number of received photons is
sufficiently large to allow the construction of time-resolved spectra, whereas
for COMPTEL and EGRET, the dearth of photons renders this detailed treatment
impossible.

In Figure 3, we present the resulting BATSE spectra for 4 different times. The
lightcurve of GRB 930131, as amply documented in the literature (Kouveliotou et
al. 1994; Ryan et al. 1994; Sommer et al. 1994), shows a sharp, intense first
pulse,\,lasting for $\sim$ 0.06 s after the BATSE trigger, followed by a second, less
intense and less sharp pulse, lasting from $\sim$ 0.75 s to $\sim$ 1.00 s after
the trigger. In between, the ``interpulse'' region of Figure 3, there
is significant yet faint
flux. Finally, there is again relatively little flux subsequently to the second pulse (lasting
for another 50 s). In Figure 3, the first pulse is further subdivided into the
spectrum for the time before the maximum flux is reached (0.00 - 0.03 s) and
that for the time after the maximum (0.03 - 0.06 s).

Since these time-resolved spectra cover only the low-energy range, a meaningful fit
to the Tavani-model (cf., Section 3.3.) cannot be done, since the value of the power-law
extension $\delta$ and the location of the peak energy $h\nu^{\ast}_{c}$ are mostly
constrained by the high-energy regime. Both the
spectra for the first and second pulses
are consistent, though, with the spectral fit (besides the normalization $C$) obtained
for the overall spectrum (cf., Figure 2). Consequently, there is no evidence
that the unusual flat morphology of the ``Superbowl-Burst'' spectrum is caused
by the superposition of individually strongly-peaked, time-variable spectra.

The spectrum between pulses is inconsistent with the average burst spectral
shape. The observed $\nu F_{\nu}$ is close to $\nu^{0}$ from 21~keV to 1~MeV
with no significant curvature or maximum. The extreme brightness of
GRB~930131 allows for this unique measure of the interpulse spectrum.
\placefigure{fig3}

\section{SUMMARY AND CONCLUSIONS}
After having given the relevant formulae for combining individual spectra, we
applied these methods to construct the broad-band spectrum of GRB 930131. With
appropriate deadtime corrections we first obtained the spectrum for the BATSE energy range,
which we then combine with the already published spectra from the COMPTEL and
EGRET TASC instruments. Broad-band spectra are fortunate occurences (multiple
instruments on board the GRO have to see a bright burst), available for only
a handful of bursts.

Within the general framework of an expanding relativistic
fireball, impacting on a surrounding medium (M\'{e}sz\'{a}ros \& Rees 1993), an attractive 
model for the production of the $\gamma$-ray photons is synchrotron emission from
a shocked and highly magnetized plasma (Tavani 1996a, b). 
This model is successful in fitting the strongly peaked
spectral shapes (in $\nu F_{\nu}$-space)
of the GRBs for which broad-band spectra have been obtained. Since our resulting spectrum
is so unusually flat, it poses an interesting challenge to the Tavani-model. As
described in Section 3.3., the model does fit well, although with a
value for the power-law component, which lies at the extreme end of the typically
encountered range, $3<\delta<6$. In the
BATSE energy-range, we were able to construct time-resolved spectra, which show
no evidence for significant evolution.


\acknowledgments{We thank D. Palmer for his suggestions 
concerning the severe deadtime problem 
in the BATSE data, as well as M. Kippen
and E. Schneid for their helpful discussions.
}

\vfill\eject

\clearpage
\figcaption{Constructing the combined spectrum in subbin $k$, ranging 
from $E_{k}$ to $E_{k+1}$.
Instruments $m$ and $m+1$ have overlapping energy bins ({\it heavy lines})
with subbin $k$, and are consequently contributing to the averaged
spectrum. Note how the subbins of the combined spectrum are defined
in accordance with the bin boundaries of the various instruments.
\label{fig1}}
\figcaption{Composite spectrum of GRB 930131. 
The spectrum shows
a low-energy portion which approaches a $\nu^{4/3}$ power law and a peak
$\nu F_{\nu}$ around 200 keV. The high-energy tail is remarkably flat. Solid
line: Best-fit to 
Tavani shocked synchrotron model.
\label{fig2}}
\figcaption{Spectra in the BATSE energy range for various times
during the burst. The panels correspond to the following
durations: (a) 0.00-0.03 s; (b) 0.03-0.06 s; (c) 0.06-0.75 s; (d) 0.75-1.00 s.
The spectra for the first and second pulse are consistent with the overall
spectrum of Fig. 1, whereas the interpulse spectrum is not. 
\label{fig3}}

\clearpage
\begin{deluxetable}{rrrrrrrrrrr} 
\scriptsize
\tablewidth{10.pc}
\tablecaption{Spectrum~of~GRB~930131 \label{spec}}
\tablecolumns{2}
\tablehead{
\colhead{$E_{low}$} &  
\colhead{} \\
\colhead{(keV)} &  
\colhead{$\nu F_{\nu}$} \\
\colhead{} &  
 } 
\startdata
 21........... & 0.079$\pm$0.029 \nl
 25........... & 0.070$\pm$0.018  \nl
 31........... & 0.089$\pm$0.016  \nl
 38........... & 0.112$\pm$0.020  \nl
 45........... & 0.095$\pm$0.021  \nl
 53........... & 0.121$\pm$0.022  \nl
 61........... & 0.148$\pm$0.025  \nl
 69........... & 0.130$\pm$0.028  \nl
 77........... & 0.182$\pm$0.033  \nl
 85........... & 0.177$\pm$0.017  \nl
 105.......... & 0.214$\pm$0.017  \nl
 139.......... & 0.223$\pm$0.025  \nl
 172.......... & 0.271$\pm$0.037  \nl
 207.......... & 0.224$\pm$0.031  \nl
 242.......... & 0.251$\pm$0.036  \nl
 288.......... & 0.269$\pm$0.053  \nl
 347.......... & 0.261$\pm$0.045  \nl
 435.......... & 0.224$\pm$0.054  \nl
 576.......... & 0.196$\pm$0.117  \nl
 671.......... & 0.212$\pm$0.110  \nl
 849.......... & 0.161$\pm$0.106  \nl
 1059......... & 0.224$\pm$0.034  \nl
 1258......... & 0.144$\pm$0.042  \nl
 1395......... & 0.095$\pm$0.044  \nl
 1528......... & 0.245$\pm$0.049  \nl
 1764......... & 0.153$\pm$0.033  \nl
 2262......... & 0.129$\pm$0.035  \nl
 2823......... & 0.169$\pm$0.040  \nl
 3528......... & 0.193$\pm$0.039  \nl
 4469......... & 0.115$\pm$0.039  \nl
 5880......... & 0.090$\pm$0.047  \nl
 7762......... & 0.228$\pm$0.060  \nl
 9643......... & 0.129$\pm$0.072  \nl
 11530........ & 0.252$\pm$0.056  \nl
 18100........ & 0.220$\pm$0.076  \nl
 29400........ & 0.180$\pm$0.114  \nl
 59510........ & 0.754$\pm$0.219  \nl
 119700....... &   \nl
 120900....... &$<$ 0.425 $\mbox{\ \ \ \ \ \ \ \ }$ \nl
 241300....... &   \nl
\enddata
\end{deluxetable}

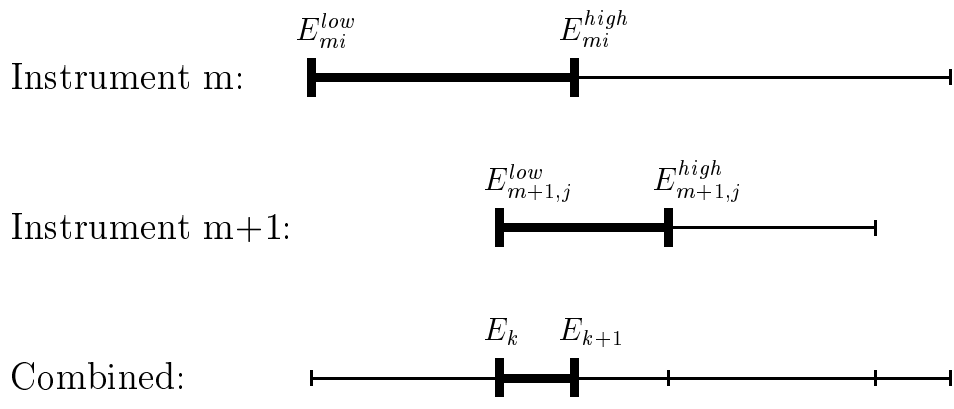
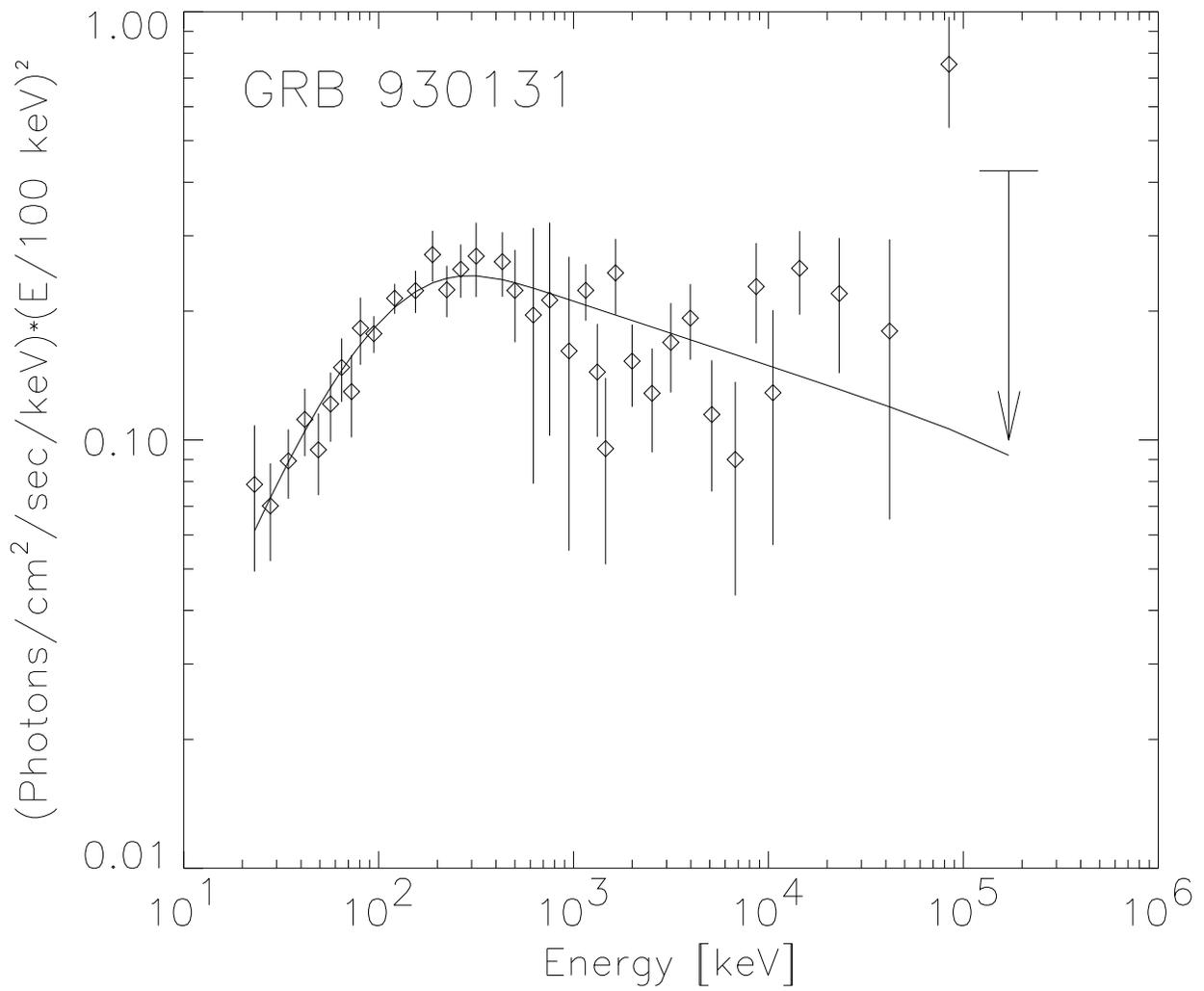
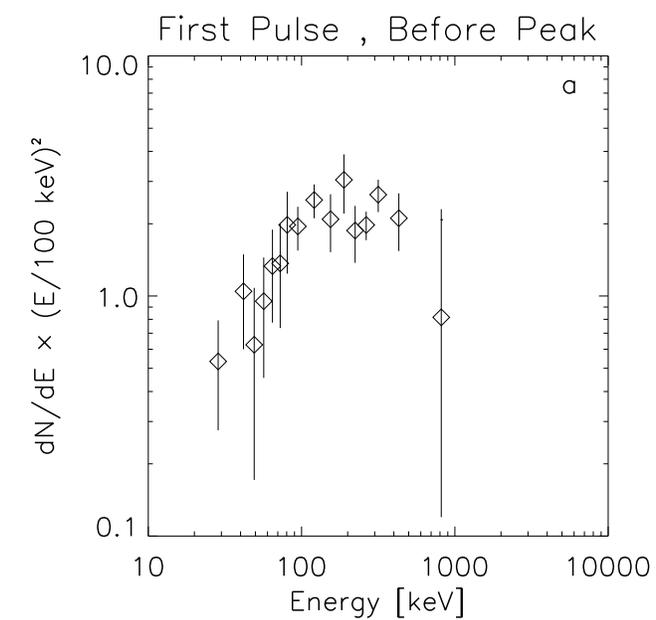
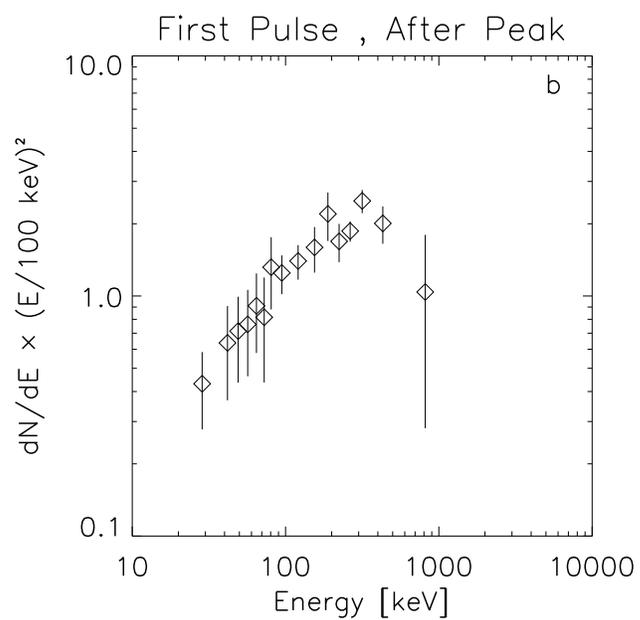
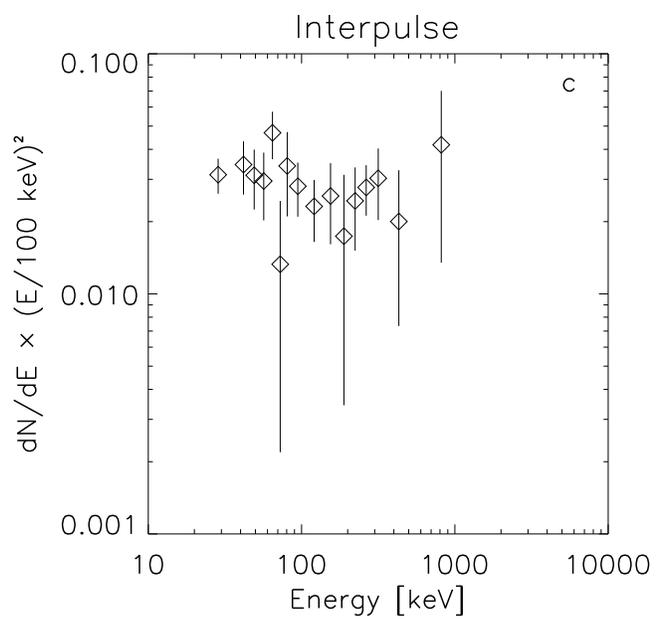
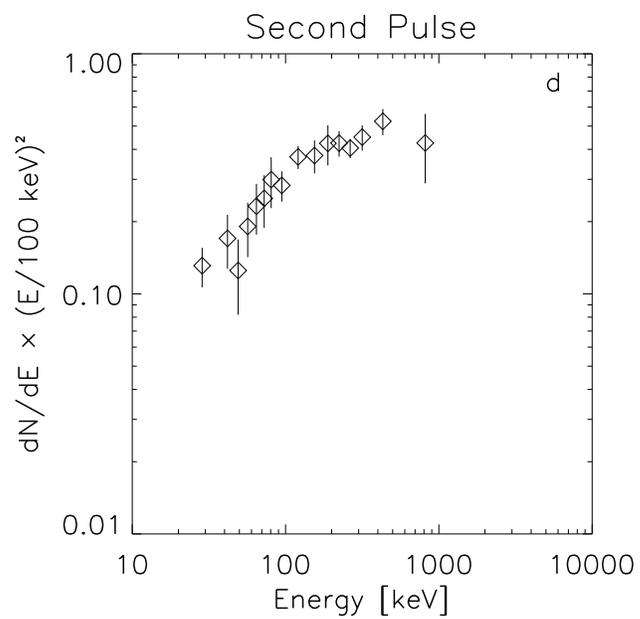
\end{document}